\documentstyle[12pt,psfig]{article}
        \def\be{\begin{equation}}
        \def\ee{\end{equation}}

\begin{document}
\begin{titlepage}

\begin{center} {\large \bf Multi-Species Asymmetric Exclusion Process in  
Ordered Sequential Update } \\

\vskip 1cm

\centerline {\bf M.E.Fouladvand$ ^{a,b}$ \footnote
{e-mail:foolad@netware2.ipm.ac.ir} ,
F.Jafarpour$ ^{a,b}$ \footnote {e-
mail:jafar@theory.ipm.ac.ir}} \vskip 1cm
{\it $^a$ Department of Physics, Sharif University of Technology, }\\
{\it P.O.Box 11365-9161, Tehran, Iran }\\
{\it $^b$ Institute for Studies in Theoretical Physics and Mathematics,}\\
{\it P.O.Box 19395-5531, Tehran, Iran}
\end{center}

\begin{abstract}

A multi-species generalization of the asymmetric simple exclusion process 
(ASEP) is studied in ordered sequential and sublattice parallel updating 
schemes. 
In this model, particles hop with their own specific  
probabilities to their rightmost empty site and fast particles overtake 
slow ones with a definite probability. Using Matrix Product Ansatz 
(MPA), we obtain the relevant algebra, and study the 
uncorrelated stationary state of the model both for an open system and on a 
ring.
A complete comparison between the physical results in these updates and 
those of random sequential introduced in [20,21] is made.\\
\end{abstract}



\end{titlepage}

\vspace {1 cm}
\newpage
\section{Introduction}
One dimensional models of particles hopping in a preferred direction 
provide simple
nontrivial realizations of systems out of thermal equilibrium [1,2,3,4]. 
In the past few
years these systems have been extensively studied and now there is 
a relatively rich
amount of results, both analytical and numerical, in the literature, 
(see [1,4] and references therein). These types of models which are 
examples of driven diffusive
systems, exhibit interesting cooperative phenomena such as boundary-induced 
phase
transition [5], spontaneous symmetry breaking [6,7] and single-defect 
induced phase
transitions [8,9,10,11,12] which are absent in one dimensional equilibrium 
systems.\\ 
A rather simple model which captures most of the mentioned features is the
 Asymmetric Simple Exclusion Process (ASEP) for which many analytical results
have been obtained in one dimension [1,4,13]. Besides its usefulness in 
describing various problems
such as kinetic of biopolymerization , surface growth , 
Burgers equation and
many others (see [4] and references therein), ASEP has a natural interpretation 
as a prototype model describing
traffic flow on a one-lane road and constitutes the basis for more 
sophisticated traffic flow models [14,15,16].\\ 

Derrida et al were first to apply Matrix Product Ansatz (MPA)
in ASEP with open boundaries [17]. Since then, MPA has been applied to many 
other interesting
stochastic models such as ASEP with a defect in the form of an additional 
particle
with a different hopping rate [11], the two species ASEP with oppositely 
charged
particles moving in same (opposite) directions [6,12,18] and many others. 
MPA has also been shown to be successful in describing disorderd ASEP-
like models. Evans [19] considered
 a model on a ring where each particle hops with its own specific rate
to its right empty site if it is empty and stops otherwise. This model 
shows two phases. In low densities the hopping rate of slowest 
particle determines the average
velocities of particles (phase I). When the density of particles 
exceeds a  
critical value, it is then the total density which determines the average 
velocity 
and the slowest particle looses its predominant role (phase II). In spite of 
its many nice features both theoretically - drawing a relevance between 
Bose-Einstein 
condensation and the observed phase transitions - and idealistic - a better 
modeling of a one way traffic flow - the possibility of exchanging between 
particles 
overtaking has
not been considered. This possibility is crucial in describing a more 
realistic traffic flow model.\\ 

Very recently in [20], a multi-species
generalization of ASEP has been proposed such that exchange processes 
among different
species has already been implemented. In this model, there are $p$-species of 
particles present in an open chain with injection(extraction) of each species 
at boundaries. Each
particle of $i$-type ( $ 1 \leq i \leq p $ ) hops forward with rate $v_i$ and 
can 
exchange its position with its right neighbour particle of $j$-type with rate 
$v_i-v_j$. \\
The subtractive form of exchange rates allows that only fast particles 
exchange their positions with
slow ones. As can be seen, this is a more realistic model for traffic flow 
in which
fast cars pass slow ones in a two-lane one-way road. In [20] using 
MPA, an infinite dimensional representation of the quadratic algebra is 
obtained but the form of currents and density profiles could not been obtained
by this infinite dimensional representation. Instead, the simple case of one 
dimensional representation was considered. Although restricting the algebra 
to be one dimensional, will
cause to loose all the correlations, but still many interesting features such 
as
a kind of Bose-Einstein condensation and boundary induced negative current 
[21], appear even in this simple uncorrelated case.\\ 

Most of the above mentioned models have been defined in continuous time, 
where the master equation of the stochastic process can be written as 
a schr\"odinger-
like equation for a "Hamiltonian" between nearest-neighbours [4,22]. 
In contrast, one can use discrete-time formulation of such random processes
and adopts other type of updating schemes
such as parallel, sub-parallel, forward and backward ordered sequential 
and particle ordered sequential (see [23] for a review). The MPA technique has 
been extended
to a sublattice parallel updating scheme [24,25,26] and in the case of open 
boundary conditions,
to ordered sequential scheme [27,28]. Although in traffic flow problems, 
parallel
updating is the most suitable one, only few exact results are 
known [15,29,30]. \\

In general, it's of prime interest to determine whether distinct 
updating schemes can produce different types of behaviour.
The present analytical results show that with changing the updating scheme of
the model, general features and phase structure remains the same but the 
value of critical parameters may undergo some changes.
In [23], Schreckenberg et al have considered ASEP under three basic updating 
procedures.
Similarities and differences have fully been discussed. Evans [29] has obtained 
analytical results in ordered and parallel updates for his model 
which was first 
solved in random sequential updating in [19]. He has demonstrated that the 
phase transition observed in [19] persists under parallel and ordered 
sequential updating.\\

In this paper, we aim to study the $p$-species model introduced in [20] under 
ordered sequential
update scheme and will show that the features observed in [20] are reproduced
in ordered updating as well. Our results will be reduced to those of [23] when 
we set $p$=1. \\

The organization of the paper is as follows. In section 2, we briefly
explain the $p$-species ASEP with random sequential updating and then describe 
the MPA in 
backward ordered sequential updating for $p$-species ASEP and will obtain
the related quadratic algebra. Section 3 contains the mapping of algebra of 
section
2 to that of [20] and includes the expressions for the currents and densities
of each type of particles in the MPA approach.
Section 4 is devoted to the one dimensional representation of the quadratic 
algebra
and the infinite limit ($ p \rightarrow \infty$) of the number of species. 
In this limit,
we use a continum description of current-density diagrams of the model.
In section 5, we consider the forward ordered updating and discuss the 
similarities 
and differences between forward and backward updating. In contrast to the usual 
ASEP
where particle-hole symmetry allows for a map of result between forward and 
backward
updating [23], here we don't have particle-hole symmetry and hence 
should separately
consider the forward updating. At the end of this section, we discuss 
the intimate relationship between sub-parallel scheme and ordered 
sequential [32].
Section 6 concludes the model with ordered updating on a closed ring. 
We obtain
current-density diagrams for both backward and forward updating.
The paper ends in section $7$ with some concluding remarks.

\section{The Model}

{\bf 2.1 ${\bf{p}}$-species ASEP in ordered sequential updating\\}

In this section we first briefly describe the $p$-species ASEP introduced
in [20]. This model consists of a one dimensional open chain of length L. 
There
are $p$ species of particles and each site contains one particle at most. 
The dynamics of the model is exclusive and totally 
asymmetric to right. Particles jump
to their rightmost site provided that site is empty, time is 
continuous and hopping of a particles of type
$i$  $(1 \leq i \leq p)$ occurs with the rate $v_i$. To cast a more realastic 
model 
for
describing traffic flow, there has been considered the possibility of 
exchanging
of two adjacent particles i.e. two neighbouring particles of types $(j)$ 
and $(i)$
swap their positions with rate $v_j-v_i$ , $v_j>v_i$.  This 
automatically forbids the exchange
between low-speed and high-speed particles  so it's a natural model for
a one way traffic flow where fast cars can overtake the slow ones. Denoting
an $i$-type particle by $A_i$ and a vacancy by $ \phi$, the bulk of the process 
is
defined by:

\begin{eqnarray} 
A_i \phi &\longrightarrow &\phi A_i \ \ \ {\rm with \ \ rate } 
\ \ \ v_i \ \ \ \ (i=1,...,p) \\
A_j A_i &\longrightarrow & A_i A_j \ \ \ {\rm with \ \ rate } \ \ \ 
v_j-v_i \ \ \ \ (j>i=1,...,p)
\end{eqnarray}

In order for all the rates to be positive, the range of $v_i$'s should 
be restricted as:
\be v_1\leq v_2 \leq v_3....\leq v_p  \ee
To complete the process, one should consider the possibility of injection and 
extraction of particles at left and right boundaries.
The injection (extraction) of particles of type $i$ at left (right) 
boundary occurs with the rate ${\alpha}_i$ 
(${\beta}_i$).\\
This completes the definition of the model. Denoting the probability 
that at time $t$, the system
contains particles of type $ {\tau}_i$ ($ \tau_i =0 $ refers to vacancy) at 
site $i$ 
($0 \leq {\tau}_i \leq p , 1 \leq i \leq L$)
by $P({\tau}_1,{\tau}_2,...,{\tau}_L,t)$, one can write the stationary state 
$P_{s}({\tau}_1,{\tau}_2,...,{\tau}_L)$ in form of a Matrix-Product-State 
(MPS)

\be 
P_{s}({\tau}_1,...,{\tau}_L) \sim \ \ <W \vert D_{\tau_1}...D_{\tau_L} \vert V>     
\ee 

in which $D_{\tau_i}$  $(0 \leq \tau_i \leq p)$ is an ordinary matrix to be 
satisfied
in some quadratic algebra induced by the dynamical rules of the model and the 
vectors $ |V> $ $ <W|$ (reflecting the effect of the boundaries) act in some
auxiliary space [31,32].
Denoting $D_0$ by $E$, the quadratic algebra reads [20]

\begin{eqnarray} 
D_i E = {1\over v_i} D_i + E \hskip 1cm  (1 \leq i \leq p)  \\
D_j D_i = {1 \over (v_i-v_j)} (v_i D_j - v_j D_i) \hskip 1cm 
(1 \leq i<j \leq p) 
\end{eqnarray}

The vectors $\vert V>$ and $<W \vert $ satisfy
\be D_i \vert V> = {v_i\over \beta_i} \vert V> \ee
\be < W \vert E =  < W\vert {v_i\over {p \alpha_i}} \ee

The rest of [20] concerns with obtaining densities and currents of species 
from
the above relations. We will come back to these results while comparing 
them with ours
in the comming sections. In what follows, we describe $p$-species 
model under ordered sequential
update.\\
As stated in the introduction, in ordered sequential updating, time is 
discrete 
and the
following events can happen in each time-step

\begin{eqnarray} 
A_i \phi &\longrightarrow &\phi A_i \ \ \ \ \ \ \ 
{ \rm with \ \ probability } 
\ \ v_i \ \ \ (i=1,...,p) \\
A_j A_i &\longrightarrow & A_i A_j \ \ \ \ \ \ \ \ 
{ \rm with\ \  probability}
\ \ f_{ji} \ \ \ (j>i=1,...,p)
\end{eqnarray}

We do not fix the form of $f_{ji}$'s and as will be seen, they will be fixed
later. Particles are also injected (extracted) at the first (last) site with 
the
probability ${\alpha}_i$ (${\beta}_i)$.
We denote the probability of the configuration $({\tau}_1,...,{\tau}_L )$
at N'th time-step by $P({\tau}_1,...,{\tau}_L;N)$.
We make a Hilbert space for each site of the lattice consisting of basis 
vectors
$\{ \vert \tau >,\tau=0,...,p \}$ where $\vert \tau >$ denotes that the site 
contains
a particle of type $\tau$ (vacancy is a particle of type 0).
The total Hilbert space of the chain is the tensor product of these local 
Hilbert
spaces. With these constructions, the state of the system at the N'th time-step 
is defined to be $\vert P,N>$ so that 

\be P({\tau}_1,...,{\tau}_L;N):=<{\tau}_1,...,{\tau}_L \vert P;N> \ee

In ordered sequential updating one can update the system from right to
left or from left to right. In general these two schemes do not produce
identical results, so it is necessary to consider both of them separately. 
We first 
consider updating from right to left (backward). The state of the system at  
$(j+1)$'th time-step is obtained from $j$'th time-step as follows
\be \vert P, j+1>=T_{\leftarrow} \vert P,j> \ee

where $T_{\leftarrow}$ is 

\be T_{\leftarrow}=L_1 T_{1,2} T_{2,3} \cdots T_{L-2,L-1} T_{L-1,L} R_L \ee

with

\be L_1=L \otimes 1 \otimes \cdots \otimes 1 \ \ \ , \ \ \  R_L=1 \otimes 1 
\otimes ... \otimes R \ee
\be T_{i,i+1}=1 \otimes 1 \cdots \underbrace 1_{i-1} \otimes {T} \otimes 
\underbrace 1_{i+2} 
\cdots 1 \otimes 1 \ee

According to (13), updating the state of the system in the next time-step  
consists of the $L+1$ sub-steps. First the site $L$ is updated: if it's empty
it's left unchanged, but if it contains a $j$-type particle $(1 \leq j \leq p)$
, this particle will be removed with the probability $\beta _i$ from the site
L of the chain, then the sites $L$ and $L-1$ are updated by acting $ T_{L-1,L}$ 
on
$\vert \tau_{L-1}> \otimes R \vert \tau_L >$.
The effect of $T_{L-1,L}$ is to update the site $L-1$ and $L$ according 
to the stochastic rules (9) and (10). After updating all the links from right 
to left, one finally updates the first site: if it's occupied it's left 
unchanged, 
if it's empty then a particle of type $i$ $(1 \leq i \leq p)$ is injected
with the probability $\alpha_i$.
This procedure defines one updating time-step. After many steps, one
expects the system to reach its stationary state $ \vert P_s>$ which must
not change under the action of $T_{\leftarrow}$ and therefore is an 
eigenvector of $T_{\leftarrow}$
with eigenvalue one \\

\be \vert P_s>=T_{\leftarrow} \vert P_s> \ee

The explicit form of $T$, $R$ and $L$ can be written as

\be 
T= \sum_{i=1}^{p} v_i (E_{0i} \otimes E_{i0} - E_{ii} \otimes E_{00}) +
\sum_{j>i=1}^{p} f_{ji} (E_{ij} \otimes E_{ji} - E_{jj} \otimes E_{ii}) + I
\ee

\be R=\sum_{i=1}^{p} {\beta}_i (E_{0i} - E_{ii}) + I \ee

\be L=\sum_{i=1}^{p} {\alpha}_i (E_{i0} - E_{00}) + I \ee

Here the matrices $E_{ij}$ act on the Hilbert space of one site and 
have the standard
definition $ (E_{ij})_{kl}= \delta_{ik} \delta_{jl}$.\\

{ \bf 2.2  Matrix Product Ansatz (MPA) for ordered sequential 
scheme (backward) }

In this section we introduce MPA for the $p$-species model with right 
to left
ordered sequential updating scheme. As shown by Krebs and Sandow [31], 
the stationary
state of an one dimensional stochastic process with arbitrary 
nearest-neighbour
interactions and random sequential update can always be written as matrix 
product state
(MPS) [31]. In [32] Rajewsky and Schreckenberg have genaralized this to 
ordered sequential and sub-parallel updating schemes which are 
intimately  related to each other.
Following [17,23] we demand that 

$$P_s( \tau_1,..., \tau_L) \sim \ \ <W \vert D_{\tau_1}...D_{\tau_L} \vert V> 
\ \ (0 \leq \tau_i \leq p)
$$

where the matrices $D_0,...,D_p$ and the vectors $\vert V>$
, $<W \vert$ are to be determined. Let's first write the above MPS in a more 
compact
form via introducing two column matrices $A$ and $\hat A$ \\

$$A = \left( \begin{array}{l}E \\ {D_1} \\ { D_2} \\ . \\ . \\ . \\ 
{ D_p} \end {array} \right) \ \ \ \ \ 
\hat A = \left( \begin {array}{l}{ \hat E } \\ {{ \hat D_1}}\\{{ \hat D_2}} \\
.\\.\\.\\ {{ \hat D_p}} \end {array} \right) \\$$

(elements of $A$ and $ \hat{A} $ are usual matrices) so we formally write

\be \vert P_s> = {1\over{Z_L}} << W \vert A \otimes A \otimes ... \otimes A 
\vert V >> \ee

where the normalization constant $Z_L$ is equal to  $<W|C^L|V>$  with 
$C= E+ \sum_{i=1}^p D_i$.
The bracket $<<\cdots>>$ indicates that the scalar product is taken in each 
entry of the vector $A\otimes A \cdots \otimes A$.
One can easily check that (20) is indeed stationary i.e. $T_{\leftarrow} 
\vert P_s>= \vert P_s>$
, if the following conditions hold

\begin{eqnarray} 
RA \vert V> = \hat A \vert V > , \\
T ( A \otimes \hat A )= \hat A \otimes A  , \\
<W \vert L \hat A= <W \vert A  
\end{eqnarray}

This simply means that a "defect" $\hat A$ is created in the beginning of an
update at site $j=L$, which is then transfered through the chain until it
reaches the left end where it disappears.
Equations (17-19) and (21-23) lead to the following quadratic algebra in the 
bulk :

\begin{eqnarray} 
[{{D_i}},{{\hat D_i}}]=[E,\hat E]=0 \ \ , \ \ i=1,...,p \\
(1-v_i){{D_i}} { \hat{E}}-{ \hat{D_i}}E=0 \ \ , \ \ i=1,...,p \\
E{{\hat D_i}}+v_i {D_i} \hat E =\hat E {D_i} \ \ , \ \ i=1,...,p\\
f_{ji}{D_j}{\hat D_i}+{D_i}{\hat D_j}={\hat D_i}{D_j} \ \ , \ \ j>i=1,...,p\\
(1-f_{ji}){D_j}{\hat D_i}={\hat D_j}{D_i} \ \ , \ \ i>i=1,...,p  
\end{eqnarray} 

and following relations\\
\begin{eqnarray} 
<W \vert (1- \sum _{i=1}^p \alpha_i){\hat E}=<W \vert E   \\
<W \vert (\alpha_i \hat E + {\hat D_i})=<W \vert {D_i} \ \ , \ \ i=1,...,p \\
(E+\sum_{i=1}^p \beta_i {D_i})\vert V>= \hat E \vert V> \\
(1-\beta_i){D_i}\vert V >= {\hat D_i} \vert V> \ \ , \ \ i=1,...,p 
\end{eqnarray} 

\section {Mapping of the ${\bf p}$-species Ordered Sequential Algebra onto 
Random Sequential Algebra}  

In this section we find a mapping between the algebra (24-32) and (5-8). 
This mapping for $p=1$ (usual ASEP) was first done in [33] where it was shown
that apart from some coefficians, ASEP in an open chain with either random or
ordered update, leads to the same quadratic algebra. Here we show that this
correspondence again holds for $p$-species ASEP. We first demand

\begin{eqnarray}
\hat E=E+e \ \ \ \ \ \ \ \\
{\hat D_i}= {D_i}-d_i \ \ , \ \ i=1,...,p
\end{eqnarray} 

where $e$ and $d_i$ are $c$-numbers. Putting (33,34) into (24-32) one arrives 
at\\
\begin{eqnarray}
v_i{D_i}E=(1-v_i)e{D_i}+d_iE \ \ , \ \ i=1,...,p  \\
f_{ji}{D_j}{D_i}=d_j {D_i}-d_i (1-f_{ji}){D_j} \ \ , \ \ j>i=1,...,p \\
<W \vert E=<W \vert e({1 \over \alpha}-1) \ \ \ \\ 
{D_i} \vert V>= {d_i \over  \beta_i} \vert V>  \ \ \ \ \  i=1,...,p
\end{eqnarray}  

in which  $\alpha = \sum_{i=1}^p \alpha_i$  and the following constraints 
must be satisfied\\

\be 
e=\sum_{i=1}^p d_i \ \ \ , \ \ 
\alpha_i=({ \alpha \over e})d_i \ \ \ (i=1,...,p)
\ee

One should note that as soon as  restricting the algebra (24-32) to the
conditions (33,34), the probabilities of injection are no longer free and are 
restricted by (39).
Up to now the exchange probabilities $f_{ji}$ have been free, however we have 
not yet 
checked associativity of the algebra (35,36). Demanding associativity fixes
these exchange probabilities to be

\be  f_{ji}={{v_j-v_i} \over {1- v_i}} \ \ \ \ , \ \ \ \ j>i=1,...,p  \ee

{\bf Remark}: according to the discrete-time nature of updating procedure, 
$f_{ji} 
$'s are more precisely, the conditional probabilities i.e. they express the
probability of exchanging between $j$ and $i$-type particles provided that  
 the $i$-type particle does not hop forward during the sub time-step. 
Thus  

\be
prob(\cdots A_iA_j \cdots;N+1| \cdots A_jA_i \cdots;N) \sim f_{ji} (1-v_i)=
v_j - v_i
\ee

Threfore we see that overtaking happens with a probability proportional to the
the relative speed. With this requirement (35-38) yield\\

\begin{eqnarray}
v_i {D_i} E=(1-v_i) e { D_i } +  d_i E \ \ , \ \ i=1,...,p  \ \ \ \\
{{D_j}} {{D_i}} = {1 \over {v_j-v_i} }  \{  d_j (1-v_i) {D_i}-d_i 
(1-v_j){D_j} \} 
\ \ , \ \ j>i=1,...,p \\                                                        
< W \vert E=<W \vert e({1 \over \alpha}-1) \\  
{ {D_i} } \vert V>= {{d_i} \over \beta_i} \vert V > , \ \ \ i=1,...,p
\end{eqnarray}  

(42-45) is the mapped algebra of $p$-species ASEP in backward ordered 
sequential                                                                                      
updating onto random sequential updating. It can be easily verified that 
similar
to one-species ASEP [17], any representation of the algebra are either one
or infinite dimensional. In the following $D_i$'s and $E$ are explicitly 
represented                                                                     

$$ \tilde{E} = 
\left( \begin{array}{lllllll} 
0&.&.&.&.&.&.\\
1&0&.&.&.&.&.\\
0&1&0&.&.&.&.\\
.&0&1&0&.&.&.\\
.&.&0&.&.&.&.\\
.&.&.&.&.&.&.\\
.&.&.&.&.&.&.\\ 
\end{array} \right ) $$ \\

$$ \tilde{D}_i = 
\left( \begin{array}{lllllll} 
\lambda_i &{\lambda_i (1-v_i) \over v_i} & \lambda_i {(1- v_i)^2 
\over {v_i}^2} &
\lambda_i {(1- v_i)^3 \over {v_i}^3} & .&.&.\\
0&{1\over v_i} & { ( 1-v_i) \over v_i }{ 1 \over v_i}&{ (1-v_i)^2 
\over {v_i}^2 } 
{ 1 \over v_i}&.&.\\
0&0&{1\over v_i} & { (1-v_i) \over v_i }{ 1 \over v_i}&.&.&.\\
.&.&.&.&.&.&.\\
.&.&.&.&.&.&.\\
.&.&.&.&.&.&.\\
.&.&.&.&.&.&.\\ 
\end{array} \right ) $$

with $\lambda _i = {1\over{(1+\eta)v_i -\eta}}$
where $\eta$ is a free parameter ( we have a class of representations).

Using (45), we multiply both side of (43) on $\vert V>$ and we obtain

\be 
v_j(1-\beta_i)-v_i(1-\beta_j)=\beta_j-\beta_i \ \ , \ \ j>i=1,...,p 
\ee

solving this equations yield

\be 
\beta_i=(1+\gamma)v_i-\gamma \ \ , \  \ i=1,...,p 
\ee

in which $\gamma$ is a new parameter which can be written in terms of known
quantities.

$$
\gamma = {{\sum_{i=1}^p (\beta_i - v_i)}\over {\sum_{i=1}^p v_i - p}}
$$

Requiring that all the probabilities to be positive, leads to the following 
condition on $v_i$'s

\be 
{ \gamma \over { \gamma+1}} \leq v_1 \leq v_2 ... \leq v_p \leq 1 \ \ , 
\ \ \gamma \epsilon [0, \infty [ 
\ee 

We conclude this section with formulas for the current operators. In contrast
to random sequential updating where currents are local i.e. caused by at 
most a single
hopping of particles, in the ordered sequential updating, the currents are 
highly nonlocal
which to say can have many hoping sources according to the multiplicative 
nature
of transition matrix $T_{\leftarrow}$. In ordered sequential updating the 
mean current  
in the $N$'th time-step through the site $k$ is defined by

\be 
<n_k^{(i)}>_{N+1}-<n_k^{(i)}>_{N}=<J_{k-1,k}^{(i)}>_{N}-<J_{k,k+1}^{(i)}>_{N} 
\ee

Our attention is concentrated on the stationary state so $N$ should go to 
$\infty$.
With introducing a bra vector 

$$<S \vert := \sum _{\tau_1,..., \tau_N} < \tau_1,...,\tau_L \vert$$

the l.h.s of (49) can be written as 

\be <S \vert n_k^{(i)} T_{\leftarrow} T_{\leftarrow}^N \vert P(0)>-
<S \vert n_k^{(i)}  
T_{\leftarrow}^N \vert P(0)> 
\ee

which in turn yields

\be 
<n_k^{(i)}>_{N+1}-<n_k^{(i)}>_N=<S \vert [n_k^{(i)},T_{\leftarrow}] 
\vert P_s> 
\ee 

We have used the fact that $<S \vert T_{\leftarrow}=<S \vert$ 
which is justified if $T_{\leftarrow}$ is the 
transfer matrix of a stochastic process .
Evaluating the commutator in (51), everything is expressed in 
stationary state expectation 
values of densities which using MPS
(20) would finally leads to the expression for the current of 
$i$-type particles from
the site $k-1$ to $k$

\be 
<J^{(i)}_{k-1,k}>_{\leftarrow}={ {<W \vert C^{k-2} 
J^{(i)} C^{L-k} \vert V>} \over <W \vert C^L \vert V>} 
\ee

in which
\be J^{(i)}=v_i{D_i} \hat E + \sum_{j>i}^{p} {{v_i-v_j} \over 
{1-v_j}}{D_i}{\hat D_j} \\
-\sum_{j>i}^{p} {{v_j-v_i} \over {1-v_i}}{D_j}{\hat D_i} 
\ee
and 
\be 
C=E+\sum_{i=1}^p {D_i} 
\ee
The first term in (53) is due to hopping of the $i$-type particles, 
the second term
corresponds to the exchanges between an $i$-type and all the particles with 
lower hopping probabilities than it and finally the last
term expresses the exchanging between all the particles with higher hopping 
probabilities and the $i$-type particle.

Using (33),(34) and the bulk algebra (42) and (43) one easily concludes that 

\be 
J^{(i)}=d_i C 
\ee

So the current and density of $i$-type particles through (at) 
site $k$ are respectively given by

\be 
<J^{(i)}_k>_{\leftarrow}=d_i{ <W \vert C^{L-1} \vert V> \over <W \vert C^L 
\vert V>} 
\ee

\be 
<n^{(i)}_k>_{\leftarrow}={<W \vert C^{k-1}{D_i}C^{L-k} \vert V> \over 
<W \vert C^L \vert V>} 
\ee

Therefore all the currents are proportional to the average current 
$J_{\leftarrow}$ 
, however $J_{\leftarrow}$ has a nontrivial dependence on hopping 
probabilities.  
The next section is devoted to the one dimensional representation of the 
algebra (42-45).
This case corresponds to the steady state characterized by a Bernouli measure. 
In spite of its simplicity, still some interesting features survive in a one
dimensional representation.\\

\section {One Dimensional Representation and Infinite-Species Limit}  

{\bf 4.1  One dimensional representation\\}  

The simplest representation of the algebra (42-45) is to take the dimension of 
the matrices
to be one. For later convenience, let us replace all $D_i$'s by ${{D_i}\over 
p}$ where $p$ is the number of species. Denoting ${D_i \over p}$ and $E$ by 
c-numbers, $ { {\cal{D}} _i \over p}$
and ${ \cal{E} }$ respectively, from equations (44) and (45) we have 

\be 
{\cal{D}} _i = { p d_i \over (1+\gamma)v_i-\gamma} \ \ 
, \ \ { \cal{E} }=e({1 \over \alpha} -1) 
\ee 

Putting these numbers in (42) leads to

\be 
v_i=1 \ \ or  \ \ {1 \over \alpha}-{1 \over \gamma}=1 
\ee 

The case $v_i=1$ corresponds to the ordinary 1-species ASEP which has been
extensively studied. Using (47) the second condition can be written as

\be 
(1- \alpha)(1- \bar{ \beta } ) = (1-e) 
\ee

in which

\be 
\alpha = \sum _{i=1}^p \alpha _i \ \ , \ \bar{\beta} = \sum_{i=1}^p { \beta_i 
\over p} 
\ee                                                                             

$\alpha$ is the total probability of injection of particles 
(note that $\alpha$
should be less than one ) and $ \bar{\beta}$ is the average
probability of extraction of particles. In the special case of 1-species (60)
reduces to 

$$(1- {\alpha}_{1})(1- {\beta}_{1})=1-{d_1}$$ 

Comparing this with the usual ASEP [33] in which
the condition for one dimensional representation reads to be $(1- \alpha)(1- 
\beta)=1-p$
( $p$ is the hopping probability), make us to take $e$ as the average 
probability
of hopping i.e. $e=\sum _{i=1}^p {v_i \over p}$. So a natural choice 
for $d_i$'s
would be to take them ${v_i \over p}$. In one dimensional representation, 
the hopping
probabilities are restricted to

\be 
{\alpha} \leq v_1\leq v_2 \leq v_3....\leq v_p \leq 1 
\ee 

Within one dimensional representation, the stationary state is 
uncorrelated and
is given by $\vert P_s>= \vert \rho >^{\otimes L}$ where

\be 
\vert \rho> = {1\over c} \left( \begin{array}{l}  {\cal{E}} \\ 
{{\cal{D}}_1\over p} \\ {{\cal{D}}_2\over p} \\.\\ . \\ .\\ {{\cal{D}}_p
\over p} 
\end{array}\right)\ \ \ , \ \
c =  {\cal{E}} + {1\over p}({\cal{D}}_1 + {\cal{D}}_2 ... {\cal{D}}_p) 
\equiv {\cal{E}} + {1\over p} {\cal{D}}
\ee 

The density and current of $i$-type particles are all site independent and
are respectively given by equations (57) and (56)

\be 
\rho_{\leftarrow}(\alpha,i) = {{{\cal{D}}_i\over p} \over 
{ e( {1 \over \alpha}-1)  + { {\cal{D}} \over p}}} \
\hskip 1cm  
J_{\leftarrow}(\alpha,i) = {{v_i\over p} \over
{ e( {1 \over \alpha}-1)  + { {\cal{D}} \over p}}} 
\ee 

One can define total density and the total current by summing over all kind 
of species and finds

\be 
\rho_{\leftarrow}(\alpha) = { { {\cal{D}} \over p} \over 
{ e ({ 1 \over \alpha }-1) + { {\cal{D}} \over p}}} 
\hskip 1cm 
J_{\leftarrow}(\alpha) = {e \over {e({1 \over \alpha }-1) + 
{ {\cal{D}} \over p}}} 
\ee\\

{\bf 4.2 Infinite-species limit\\}

At this stage we consider the limit $p \rightarrow \infty $, and we assume
that the hopping probabilities of particles are chosen from a continuous
distribution $P(v)$. Discrete quantities ${1 \over p} F(i)$ are transformed 
into
$f(v)P(v)$ and sums into integrals. Equations (64) and (65) take the form\\

\be  
\rho_{\leftarrow}(\alpha,v)= { {\cal{D}} (\alpha,v)P(v) \over 
{e({1 \over \alpha}-1) + {\cal{D}}(\alpha)}} \hskip 2cm  
J_{\leftarrow}(\alpha,v) = {v P(v) \over {e({1 \over \alpha}-1)  + 
{\cal{D}}(\alpha)}} \ee 

\be 
\rho_{\leftarrow}(\alpha)={{\cal{D}}(\alpha) \over e({1 \over \alpha}-1)+ 
{\cal{D}}(\alpha)}  \hskip 2cm 
J_{\leftarrow}(\alpha) = { e \over {e({1 \over \alpha}-1)  
+ {\cal{D}}(\alpha)} } 
\ee     

where

\be 
{\cal{D}}(\alpha,v) ={(1- \alpha) v P(v) \over { v-\alpha}} \ \ {\rm  and} 
\ \ {\cal{D}}(\alpha) = (1- \alpha) \int_{\alpha}^1 {v  \over { v-\alpha}} 
P(v) dv 
\ee 

Although one has many choices for $P(v)$, we first take the following [19]. 
It has the merit that ${\cal{D}}(\alpha)$ can be analytically evaluated.

\be 
P_1(v)={(m+1) \over (1- \alpha)^{m+1}}(v- \alpha)^m \ \ , \ \  {m \geq 0}  
\ee

This is a normalized distribution that vanishes with some positive power in 
low-velocities
and increases up to $v=1$. The average hopping probability $e$ is found to be 

$$
e= \int_{\alpha}^{1} v P_1(v) dv= {(m+1) \over (m+2)}(1-\alpha)+\alpha 
$$ 

expressing $m$ in terms of $e$ and $\alpha$ we have 

\be 
m={{2e-\alpha-1} \over {1-e}} 
\ee

for $m$ to be positive, (70) implies $(e,\alpha \leq 1)$

\be 
2e-\alpha-1 \geq 0 
\ee

We first study the current-density relationship for a fixed hopping 
probability,
$e$. In order to do this, we evaluate ${\cal{D}}(\alpha)$ with (68) and 
replace $m$ from (70)                                                                           

\be 
J_{\leftarrow}(\alpha,e)={ e \over { e ({1 \over \alpha}-1) + 
{ {2e-1- \alpha e} \over {2e-1- \alpha}} }}
\ee

\be
\rho_{\leftarrow}(\alpha,e)= { { {2e-1- \alpha e} \over {2e-1- \alpha} } \over 
{ e( {1 \over \alpha}-1) + {{2e-1- \alpha e} \over {2e-1- \alpha}}}} 
\ee

The above expressions gives the total current and total density in terms of
two control parameters namely the total arrival probability $\alpha$ and the
average hopping probabilty $e$.\\
We now eliminate $\alpha$ between $J_{\leftarrow}$ and $ \rho_{\leftarrow}$ 
numerically which then gives the current density diagram. This diagram is 
shown in figure (1) for two values of $e$. \\

\begin{figure}
\centerline{\psfig{figure=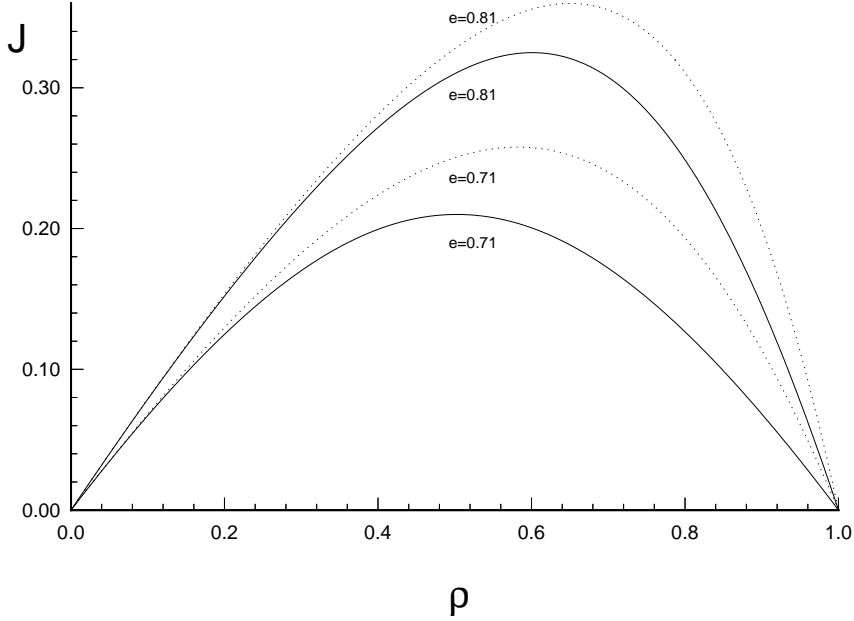,width=.8\columnwidth}} 
\vspace{5 mm}
\caption{ The current versus the density for different values of $e$ in 
backward updating. Continuous lines refer to $P_{1}(v)$ and 
dotted lines refer to $P_{2}(v)$.}
\end{figure}

{\bf Remark}: 
Total current $J_{\leftarrow}$ and total density $\rho_{\leftarrow}$ are in 
general functions of three control parameters $e$, $\alpha$ and $m$. Recalling
that $e$ is the average hopping probabilty , $\alpha$ is the total rate of
injection and $m$ determines the shape of hopping distribution function. 
Equation (70) implies that only two parameters are independent. There is a 
one-to-one                                                                                                      
correspondence between the two dimensional parameter space defined by the 
surface (70) and the current-density
space. $J_{\leftarrow}$ versus $\rho_{\leftarrow}$ in fig (1) corresponds to
ntersection of planes $e=$ constant , with the surface defined by (70). We can 
instead
look at the intersection of $\alpha =$ constant planes with the surface and 
find 
the corresponding curves in $J_{\leftarrow}-\rho_{\leftarrow}$ plane. 
This is done
by eliminating $e$ between equations (72) and (73). Figure (2) shows these 
diagrams for some values of $\alpha$. 

\begin{figure}
\centerline{\psfig{figure=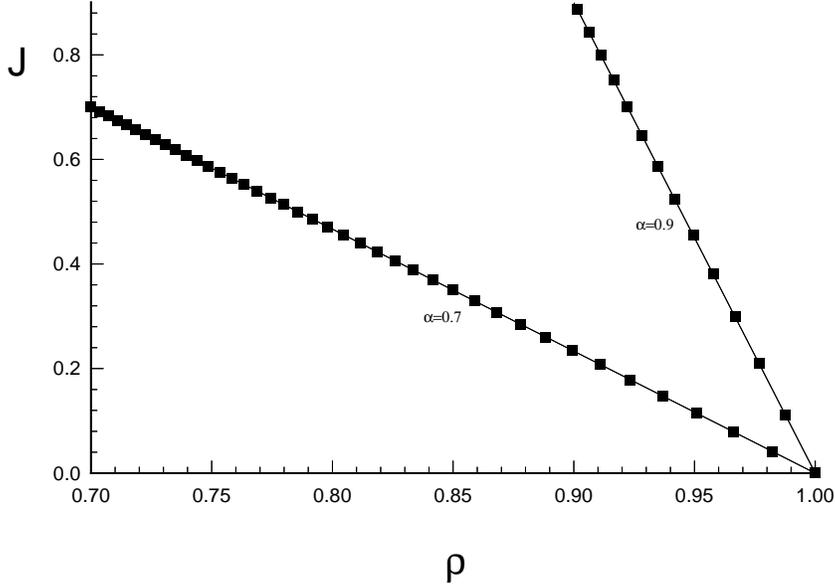,width=.8\columnwidth}} 
\vspace{5 mm}
\caption{ The current versus the density for different values of $\alpha$ in 
backward updating. Continuous lines refer to $P_{1}(v)$ and filled squares 
refer to $P_{2}(v)$.}

\end{figure}

Finally we consider the curves of constant $m$ in $J_{\leftarrow}-
\rho_{\leftarrow}$ plane.
To obtain these curves, one should write $J_{\leftarrow}$ and 
$\rho_{\leftarrow}$ in terms
of $\alpha$ and $m$ as follows

\be
J_{\leftarrow}(\alpha,m)= {{\alpha(\alpha + m +1)}\over{(\alpha + m +1)(1-
\alpha) + \alpha (m+2)(1+{{\alpha}\over m})}}
\ee

\be
\rho_{\leftarrow}(\alpha,m)= {{\alpha(m +2)(1+{{\alpha}\over m})}\over
{(\alpha + m +1)(1-
\alpha) + \alpha (m+2)(1+{{\alpha}\over m})}}
\ee

Eliminating $\alpha$ between $\rho_{\leftarrow}(\alpha,m)$ and $J_{\leftarrow
}(\alpha,m)$ would give us the current-density diagrams for a fixed value of
$m$. Figure (3) shows these diagrams for some values of $m$. As can be seen, 
the
current does not vanish at $\rho_{\leftarrow} =1$. This can be explained by 
noticing
that although at $\rho_{\leftarrow} =1$, the chain is completely filled, still 
we have current via exchange processes. 
At $ \rho_{\leftarrow}=1 $, the more decreasing $m$, the more $J_{\leftarrow}$ 
approaches to zero.

\begin{figure}
\centerline{\psfig{figure=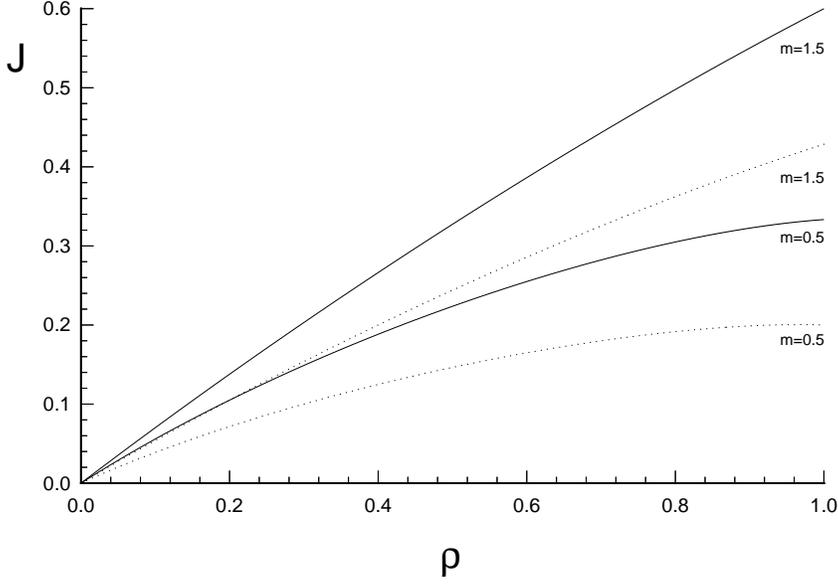,width=.8\columnwidth}} 
\vspace{5 mm}
\caption{ The current versus the density for different values of $m$ in 
backward updating. Continuous lines refer to $P_{1}(v)$ and dotted lines 
refer to $P_{2}(v)$.}
\end{figure}

Using (72) and (74), we can also look at the behaviour of current itself as a 
function of control parameters. In figures (4) and (5), we show the dependence
of $J_{\leftarrow}$ on $\alpha$ ,$e$ for some fixed values of $e$ and $\alpha$.
Note that for each $\alpha$, there is 
a lower limit of $e$ which can be obtained through equation (70). \\

\begin{figure}
\centerline{\psfig{figure=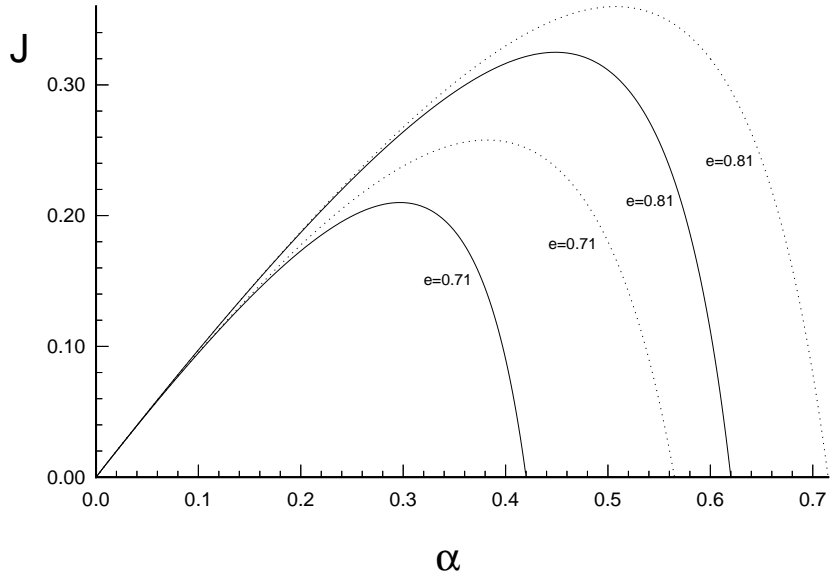,width=.8\columnwidth}} 
\vspace{5 mm}
\caption{ The current versus the arrival probability of particles for 
different values of $e$ in backward updating. Continuous lines refer to 
$P_{1}(v)$ and dotted lines refer to $P_{2}(v)$.}
\end{figure}

\begin{figure}
\centerline{\psfig{figure=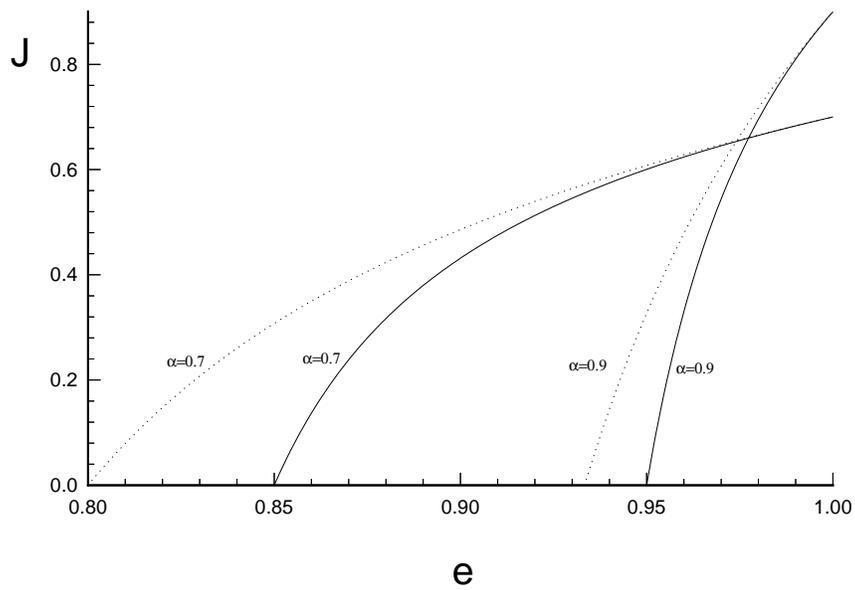,width=.8\columnwidth}} 
\vspace{5 mm}
\caption{ The current versus the total probability of hopping for 
different values of $\alpha$ in backward updating. Continuous lines refer 
to $P_{1}(v)$ and dotted lines refer to $P_{2}(v)$.}
\end{figure}

\newpage

Our second choice of velocity distribution function is the following

\be
P_2(v)= {{(m+1)(m+2)}\over {(1-\alpha)^{m+2}}} (v-\alpha)^m(1-v) \ \ \  m\geq 
0
\ee

It vanishes at $v=\alpha$ , $v=1$ and has a maximum at $v_{max}={{m+\alpha}
\over{
m+1}}$. If $m$ increases, $v_{max}$ approaches to one and if $m$ decreases to 
zero , it approaches to $ \alpha $. Inserting $P_2(v)$ into (39) we  
arrive at 

\be
m = {{3e-2\alpha -1}\over{1-e}}
\ee

using (67),(68) and (77), we express $J_{\leftarrow}$ and $ \rho_{\leftarrow}$ 
in terms of $e$, $\alpha$ and $\alpha$, $m$

\be
J_{\leftarrow}(\alpha,e) = {{e\alpha(2\alpha + 1-3e)}\over{e(1- \alpha)
(2\alpha -3e +1)
+ \alpha(2\alpha e -3e +1)}}
\ee

\be
\rho_{\leftarrow}(\alpha,e) = {{\alpha(2\alpha e+ 1-3e)}\over{e(1- \alpha)
(2\alpha -3e +1)
+ \alpha(2\alpha e -3e +1)}}
\ee

\be
J_{\leftarrow}(\alpha,m) = {{\alpha(2\alpha + m+1)}\over{(1- \alpha)
(2\alpha +m +1)
+ \alpha(m +3)({{2\alpha}\over m} +1) }}
\ee

\be
\rho_{\leftarrow}(\alpha,m) = {{\alpha(m+3)({{2\alpha}\over m} +1) }
\over{(1- \alpha)(2\alpha +m +1)
+ \alpha(m +3)({{2\alpha}\over m} +1) }}
\ee

We now eliminate $\alpha$ between $J_{\leftarrow}(\alpha,e)$ and 
$\rho_{\leftarrow}(\alpha,e)$
which leads to current-density diagrams for fixed values of $e$. 
Dotted lines in fig (1) shows
these diagrams for the same values of $e$ .\\
Similar to $P_1(v)$, we can consider the current-density diagrams 
corresponding
to constant $\alpha$ and $m$. These diagrams are shown by dotted 
lines in figures
(2) and (3) respectively. Dependence of $J_{\leftarrow}$ on $\alpha$ 
and $e$ 
for $P_2(v)$ are also shown in figures (4) and (5) by dotted lines. 
Note that in figure (5), the curves obtained from $P_{1}(v)$ 
asymptotically approach to
those of $P_{2}(v)$.\\
Here, we would like to reveal a feature of the infinite species limit which
is somehow reminiscent of Bose-Einestein condensation [19]. 
Equation (68) implies that the density
of particles with speed $v$ is proportional to ${ vP(v) \over {v-\alpha}}$.
Taking (69,76) for $P(v)$ we have 

\be 
\rho(v) \sim v(v- \alpha )^{m-1} 
\ee

Recalling that $\alpha$ is the minimum speed of particles, equation (82) shows
two different kinds of behaviour depending on whether $m>1$ or $m<1$.\\

I) If $m-1>0$ then $\rho(v) \rightarrow 0 $ for $ v \rightarrow \alpha $ \\

which means that density of low speed particles is small, i.e. most of the 
particles
move with rather high speed.\\

II) If $m-1<0$ then $\rho(v) \rightarrow \infty $ for $ v \rightarrow \alpha $.
\\

In contrast to the case I, here the density of low-speed particles are large
and most of the particles move with low speed , which can be interpreted 
as appearing of the traffic jammed phase. \\

\section { ${\bf{p}}$-Spesies ASEP with Forward Updating }

{\bf 5.1  Formulation } \\ 

As stated in the introduction and section (2), instead of right to left
(backward) updating, one can change the direction of updating and starts from
the first site of the chain (forward updating) and updates from the left to 
the right 
in the same manner of backward updating. Most of the steps are similar 
to backward updating and we
only write the results. The transfer matrix takes the following form 

\be 
T_{\rightarrow}=R_L T_{L-1,L}...T_{1,2} L_1 
\ee 

All the matrices are the same as in (17,18,19). The MPS for the steady state
is written as [23]

\be 
\vert P_s>_{\rightarrow}=<< W \vert \hat A \otimes \hat A \otimes ... \otimes 
\hat A \vert V >> 
\ee

Taking $A$ and $ \hat A$ to satisfy the same algebra (21-23), 
makes $\vert P_s>_{\rightarrow}$
to be a stationary state i.e. $T_{\rightarrow} 
\vert P_s>_{\rightarrow} = \vert P_s>_{\rightarrow}$.
Here at first site $i=1$ a "defect" $A$ is created, then transmited 
forward until
it reaches the last site $i=L$ where it disappears.
Next we consider formulas for the currents and densities. 
Here the situation is quite
different and the difference between forward and backward updating 
reveals itself.
The definition of currents reads from (49-51) and $T_{\leftarrow}$ 
is replaced
with (83). The mean current of $i$-type particles through site $k$ is 
found to be \\
\be <J_{k-1,k}^{(i)}>_{\rightarrow}={ {<W \vert \hat{C}^{k-2} J^{(i)} 
\hat{C}^{L-k} \vert V > } \over { <W \vert \hat C^L \vert V > } } \ee \\
Where $J^{(i)}$ is the same as equation (53), and $ \hat C =\hat E+ 
\sum_{i=1}^{p}
{\hat D_i } $. 
We again demand that $\hat E$ and $ { \hat D_i }$ satisfy equation (33,34)
which in turn let us revisit equation (42-45) and thus we have 

\be 
J^{(i)} = d_i C 
\ee

\be
\hat{C} =C  
\ee

Putting (86,87) in (85) yields\\

\be  
<J^{(i)}>_{\rightarrow}=d_i { { < W \vert C^{L-2} \vert V > } 
\over { <W \vert {C}^L \vert V > } } 
\ee 

Also one can write the mean density of $i$-type particles at site 
$k$ \\

\be 
<n_{k}^{(i)}>_{\rightarrow}= { < W \vert {C}^{k-1} ( D_i-v_i )  
C^{L-k} \vert V >  
\over 
 <W \vert C^{L} \vert V >  } 
\ee \\

{\bf 5.2  One dimensional representation and infinite number of species 
limit in forward updating \\ }

Again scaling all $D_i$'s by a ${1\over p}$ factor, we now take ${ D_i 
\over p}$ 
and $E$ to be c-numbers. 
Similar to backward update, they are ${{\cal{D}}_i \over p} $
and ${\cal{E}}$ respectively and the equations (58-61) remain the same.
In one dimensional representation, the densities and the currents of $i$-type 
particles are all site independent and are respectively given by

\be 
\rho_{\rightarrow}(\alpha,i) = { ({{\cal{D}}_i\over p}-{v_i \over p}) 
\over {e({1 \over \alpha}-1) + {{\cal{D}} \over p}}} \
\ \ , \ \ J_{\rightarrow}(\alpha,i) = {{v_i\over p} \over
{e({1 \over \alpha}-1) + { {\cal{D}} \over p}}} 
\ee 

Comparing the above equations with their counterparts in backward updating, 
we see that
currents do not change but forward density undergoes the following modification

\be 
J_{\rightarrow}( \alpha,i)= J_{\leftarrow}( \alpha,i)= J( \alpha,i) \ \ , \ \  
 \rho_{\rightarrow}( \alpha,i)= \rho_{\leftarrow}( \alpha,i)-J(\alpha,i) 
\ee  

The above relations reveals the difference between forward and backward
updating. Similar relation between backward and forward densities is seen in 
[23].
We again define the total density and current by summing over 
densities and currents of all kind of species\\

\be 
J_{\rightarrow}( \alpha ) = J_{\leftarrow}( \alpha )= 
{ e \over { e({1 \over \alpha}-1)+ { {\cal{D}} \over p} }} \ \ , \ \
\rho_{\rightarrow}( \alpha ) = \rho_{\leftarrow}( \alpha ) - J( \alpha) 
\ee  

Now we take the limit of $p \rightarrow  \infty$. Adopting the same
distribution functions $P_1(v)$ , $P_2(v)$ and using (92), one easily 
can obtain $J_{\rightarrow}$ and $\rho_{\rightarrow}$ as functions of $e$ , 
$\alpha$ and $m$ , both for $P_1(v)$ and $P_2(v)$. Similar to the backward 
scheme, the corresponding current-density diagrams can be obtained by 
eliminating one of the control parameters. These diagrams are shown in 
figures (6) to (8).

\begin{figure}
\centerline{\psfig{figure=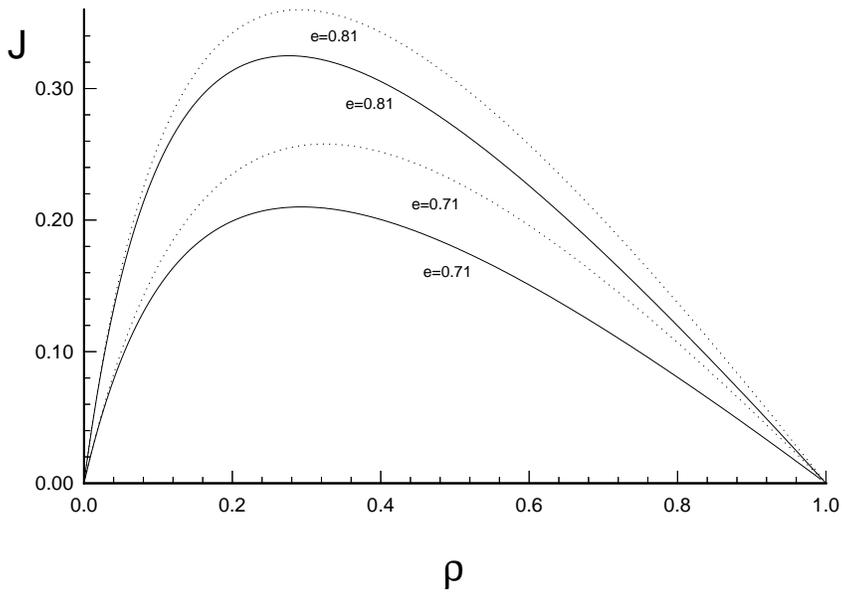,width=.8\columnwidth}} 
\vspace{5 mm}
\caption{ The current versus the density for different values of $e$ 
in forward updating. Continuous lines refer to $P_{1}(v)$ and dotted 
lines refer to $P_{2}(v)$.}
\end{figure}

\begin{figure}
\centerline{\psfig{figure=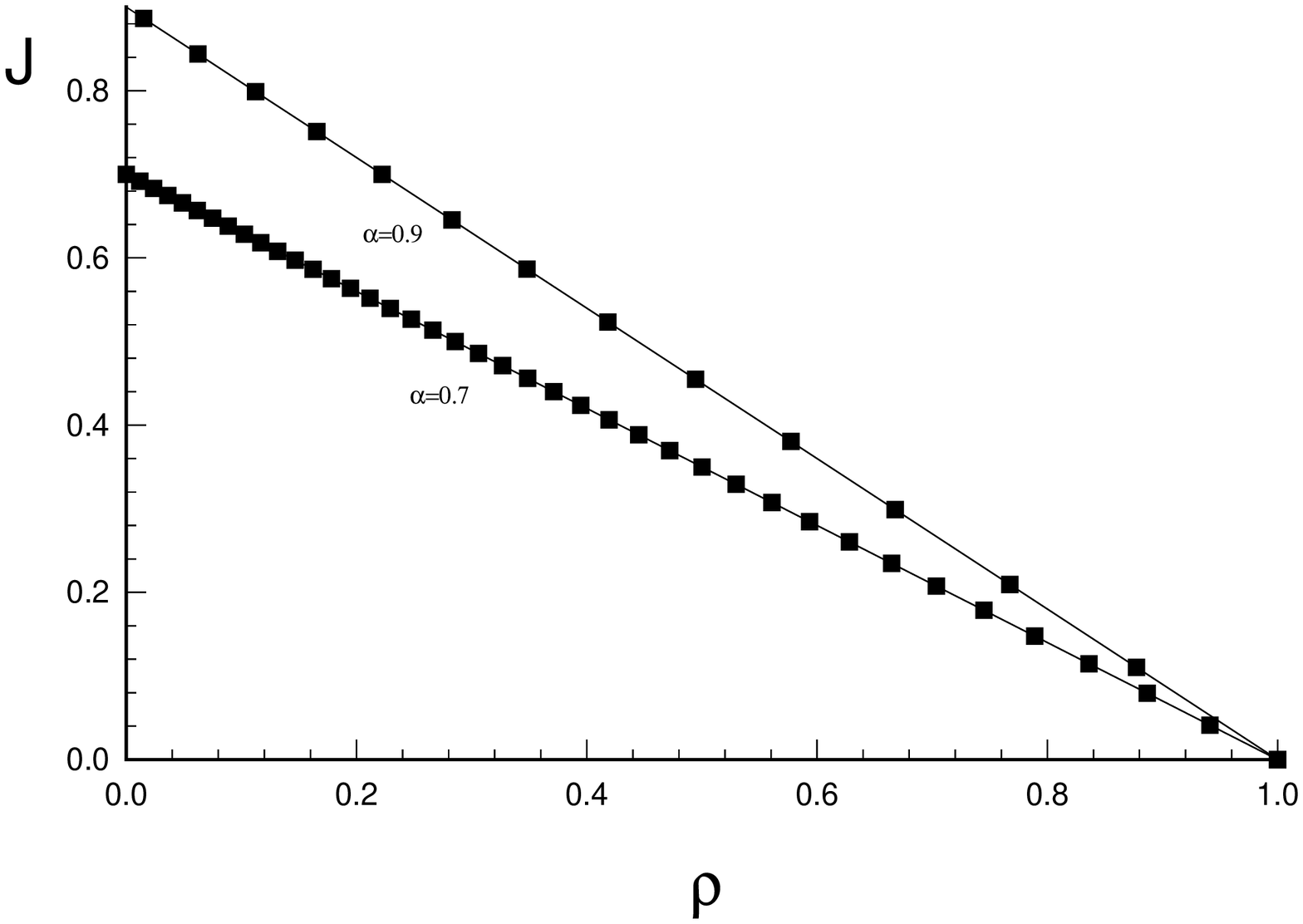,width=.8\columnwidth}} 
\vspace{5 mm}
\caption{ The current versus the density for different values of $\alpha$ 
in forward updating. Continuous lines refer to $P_{1}(v)$ and filled squares 
refer to $P_{2}(v)$.}
\end{figure}

\begin{figure}
\centerline{\psfig{figure=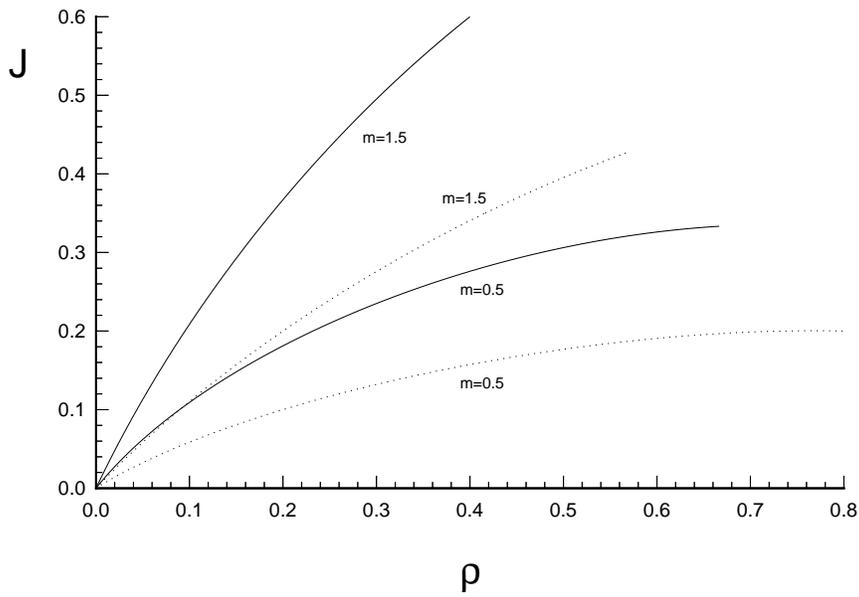,width=.8\columnwidth}} 
\vspace{5 mm}
\caption{ The current versus the density for different values of $m$ 
in forward updating. Continuous lines refer to $P_{1}(v)$ and dotted 
lines refer to $P_{2}(v)$.}
\end{figure}

\newpage

{\bf Remark}:\\
Surprisingly as can be seen in fig(7), when $\rho_{\rightarrow}$ goes to zero,
the value of $J_{\rightarrow}$ does not vanish. This is in contrast to other's 
results and is an exclusive effect appearing only in forward updating which
can be explained by noting that, when the lattice is completely empty, in
first site a particle is injected with the probability $\alpha$ and according 
to the multiplicative nature of
the transition matrix is transfered through the lattice, hence one has a 
non-zero current. \\
In general, the value of $ J_{\rightarrow} $ at $ \rho_{\rightarrow}=0$ is
equal to $\alpha$ and this point refers to the point $(m= \infty, e=1, \alpha)$
in parameter space.\\

We would like to end this section with some remarks on sub-parallel updating  
scheme. In fact as stated in section 1, there are few exact results in parallel 
updating. The root of this difficulty is the non-local nature of transfer 
matrix
which in contrast to the ordered sequential updating, can not be written as a 
product
of local transfer matrices. A simpler case is to consider a sub-parallel 
updating
scheme. In this scheme, one proceeds with two half time-steps. In the first 
half,
one updates the first site, last site and all pairs $(\tau_i, \tau_{i+1})$ 
with 
an even $i$ ($L$ is taken to be even). Then in the second half time-step, 
one 
updates all pairs $(\tau_i, \tau_{i+1})$ with $i$ odd. So the transfer matrix 
is

\be
T_{sp}=T_{sp}^{(2)} T_{sp}^{(1)}
\ee

with

\be
T_{sp}^{(1)} = L_1 T_{2,3} T_{4,5} \cdots T_{L-2,L-1} R_L
\ee

\be
T_{sp}^{(2)} =  T_{1,2} T_{3,4} \cdots T_{L-1,L}     
\ee 

Defining MPS for sub-parallel updating as follows [25]

\be
|P_s>_{sp} = <<W|\hat {A} \otimes A \otimes \hat {A} \otimes \cdots \hat {A} 
\otimes
A|V>>
\ee 

It can be verified that $ T_{sp} |P_s>_{sp} = |P_s>_{sp}$ provided that 
equations (21-23) are satisfied.

It is shown in [32] that sub-parallel and ordered sequential
updating schemes are intimately related to each other. It is proved that in 
general the following correspondence exists

\be 
<n_{k}^{(i)} >_{sp}= 
\cases{ <n_{k}^{(i)} >_{\rightarrow} &  $k$  odd
\cr <n_{k}^{(i)}>_{\leftarrow} & $k$  even \cr} 
\ee
                                       
\be 
<n_{k}^{(i)} n_{l}^{ (j)} >_{sp}= 
\cases{ <n_{k}^{(i)} n_{l}^{ (j)} >_{\rightarrow} &  $k$ , $l$ odd
\cr <n_{k}^{(i)} n_{l}^{ (j)} >_{\leftarrow} & $k$ , $l$ even \cr} 
\ee

where $k$ and $l$ refer to the lattice sites and $i$ and $j$
refer to the state of the site.
Using this general correspondence, we obtain the density profile of 
$p$-species
ASEP under sub-parallel updating ( one dimensional representation )

\be
<n_{k}^{(i)}>_{sp}={ {{\cal{D}}_i \over p} \over e( {1 \over \alpha} -1) + 
{ {\cal{D}} \over p}} \ \  \ \ k=even
\ee

\be
<n_{k}^{(i)}>_{sp}={ {{\cal{D}}_i \over p}-{v_i \over p} \over 
e( {1 \over \alpha}                                                             
-1) + { {\cal{D}} \over p}} \ \  \ \ k=odd
\ee

\section{ ${\bf p } $-Species ASEP with Ordered Updating on a Ring}

In this section we consider the $p$-species ASEP on a closed ring of $N$ sites.
 We work in a canonical ensemble in which the number of each species $(i)$ is 
fixed to be $m_i$ and we take the total number of particles to be $M$ i.e. 
$\sum_
{i=1}^p m_i = M$ . \\
The periodic system can be described by a one dimensional
representation of the bulk algebra (24-28). In this case
the bulk algebra reduces to the following equations

\be
(1-v_i) d_i \hat{e} = \hat{d}_i e
\ee
\be
({ {1- v_j } \over {1-v_i}}) d_j \hat {d_i} = \hat {d_j} d_i
\ee

The above equations yield

\be
\hat {d_j} = {{\hat {e}}\over {e}} (1-v_j)d_j
\ee

Here $d_i$ and $\hat{d_i}$, correspond to one dimensional representations
of $D_i$ and $\hat{D_i}$  (not to be confused with those introduced in (34)).
Using (53) and (57) we obtain 
the following forms for the density and the current of $i$-type particles : 

\be  
\rho_{\leftarrow}^{(i)} = { {d_i} \over {e+ \sum_i d_i}} , \ \ \ 
J_{\leftarrow}^{(i)} =  \hat {e}(v_id_i + {1\over{e}} [v_id_i \sum_j d_j - d_i 
\sum_j d_jv_j]) 
\ee                             

Summing over $i$, we obtain the total current and density

\be
\rho_{\leftarrow} = {  \sum_i d_i \over {e+ \sum_i d_i} }   
\ \ , \ \ 
J_{\leftarrow}=
\hat{e} \sum_i {v_id_i}
\ee                                

Defining the population averaged velocity ${<v>}$ as follows

\be
<v>=
{{\sum_i m_iv_i}\over { \sum_i m_i}}
\ee

and rescaling the $d_i$'s and $e$ so that

\be
e+ \sum_i d_i = \hat{e} + \sum_i \hat{d_i} = 1
\ee

we arrive at 

\be
J_{\leftarrow} = {{ {<v>} \rho_{\leftarrow} (1 - \rho_{\leftarrow})} \over
{ 1 - {<v>} \rho_{\leftarrow} }}
\ee   

which is the current-density relation of $p$-species ASEP on a ring with 
backward
updating. Comparing it with the usual ASEP on ring with backward updating in
[23] , we see that they both have the same form. In $p$ species model, 
${<v>}$ plays the role of hopping probability in usual ASEP. Fig (9) shows
$J_{\leftarrow}$ versus $ \rho_{\leftarrow} $ for different values of ${<v>}$.

\begin{figure}
\centerline{\psfig{figure=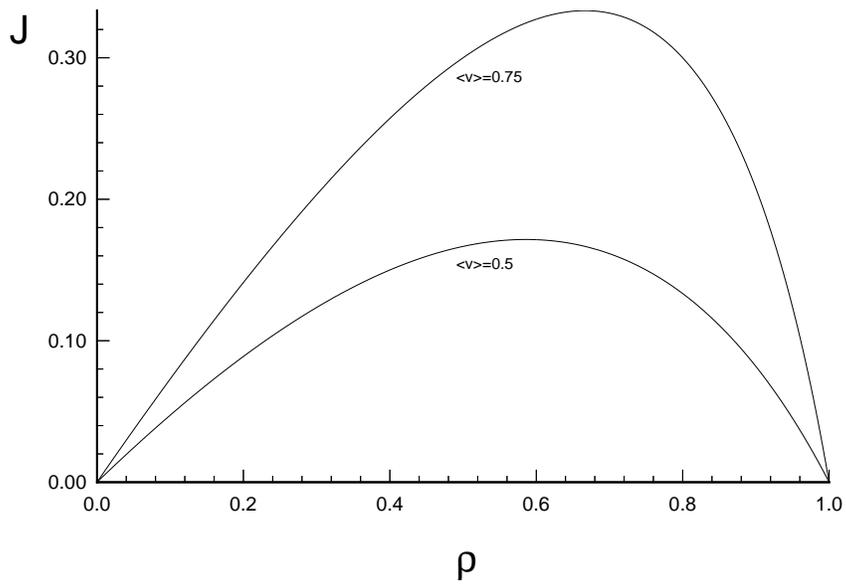,width=.8\columnwidth}} 
\vspace{5 mm}
\caption{ The current versus the density for different values of $<v>$ 
in backward updating. }
\end{figure}

The maximum current occurs at

\be
\rho_{ \leftarrow }^{max} ( {<v>} ) = { 1 - (1-( {<v>})^{ 1 \over 2} )  
\over  {<v>} } \geq {1\over 2} 
\ee

We now consider the forward updating. Note that since we don't have 
particle-hole   
symmetry, the current-density relation in forward updating can not be obtain
from the one in backward updating and should be considered seperately. 
In forward
updating we have

\be
\rho_{\rightarrow}^{(i)}={ \hat{d_i} \over \hat{e}+\sum_i \hat{d_i}} \ \ , 
\ \ 
J_{\leftarrow}^{(i)}=J_{\rightarrow}^{(i)}
\ee
                                                                                 
Using (101-103) and (107), after straightforward calculations, we 
arrive at 

\be
J_{\rightarrow}={ (1- \rho_{\rightarrow}) \rho_{\rightarrow} <{v \over 1-v}> 
\over
1+ \rho_{\rightarrow}<{v \over 1-v}>}
\ee

where                                                                           

\be                 
<{v \over 1-v}>={ \sum_i {v_i \over 1-v_i} m_i  \over  \sum_i m_i}
\ee
\\
If we now take $p=1$ , $ <{v \over 1-v}>$ will reduce to  $v_1 \over 1-v_1 $
and (111) takes the following form

\be
J_{\rightarrow} = {{ v_1 \rho_{\rightarrow} (1 - \rho_{\rightarrow})} \over
{ 1 - {v_1} \rho_{\rightarrow} }}
\ee
\\
and the particle-hole symmetry is recovered [23] i.e. (113) is obtained from 
(108) by changing $ \rho_{\leftarrow}$ to $1- \rho_{\rightarrow}$.\\

Fig(10) shows $J_{\rightarrow}$ versus $\rho_{\rightarrow}$ for different 
values of $<{ v \over 1-v}>$.

The maximum of $J_{\rightarrow}$ has moved to the left. This maximum occurs at 

\be
\rho_{\rightarrow}^{max}(< {v \over 1-v}>)={ 1 \over <{ v \over 1-v}>}
[ (1+ < {v \over 1-v} > )^{1 \over 2} -1 ] \leq {1 \over 2} 
\ee

\newpage

\begin{figure}
\centerline{\psfig{figure=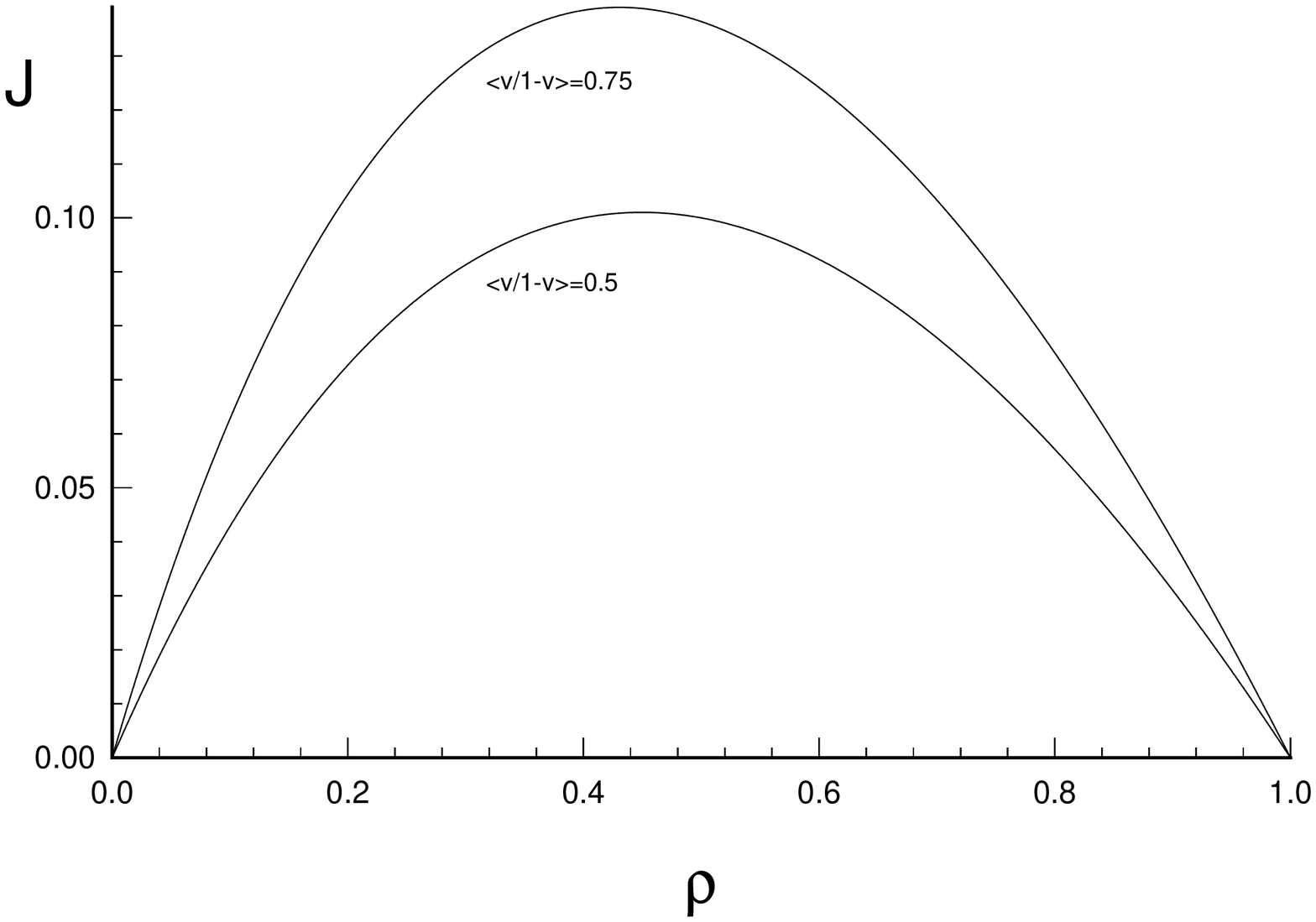,width=.8\columnwidth}} 
\vspace{5 mm}
\caption{ The current versus the density for 
different values of $<{v \over {1-v}}>$ in backward updating. }
\end{figure}

\section{Comparison and Concluding Remarks}

Here we compare our results with those of [20] and specify the similarities
and differences between ordered and random sequential updating procedures. 
Both
procedures are described by a similar quadratic algebra. In random update, 
time
should be so rescaled such that the average hopping rate equals one. 
On the contrary in ordered updating 
the average hopping probability $e$ remains as a free parameter. 
This is one of
the main differences between two updating schemes. In both schemes, 
injection rate (probability) of an $i$-type particle ($\alpha_i$) is 
proportional
to its hopping rate (probability) $v_i$.\\
When considering infinite species limit, one can investigate the 
characteristics
of both schemes with a limited number of control parameters. As long as 
analytical
calculations are concerned, these control parameters are $\alpha$ , 
$e$ and $m$ 
in ordered schemes and $\alpha$, $m$ and $\lambda$ in random scheme 
where $m$, 
$\lambda$
determine the shape of distribution function [20]. One of the advantages of 
ordered scheme is the appearance of the more physical parameter $e$ in 
control
parameters, which is absent in random scheme. \\
In this paper, we made a more complete investigation of the current-
density and current diagrams for different regions of parameter space. \\
We also evaluated the dependence of the current on the density for fixed 
values of
$\alpha$ in random scheme.
The corresponding diagram is quite similar to ours in 
figure (2) but the values of current and minimum allowed value of the density
are different.\\ 
All the result of this paper and [20] have been obtained in a 
restricted region of parameters space ($\alpha_i, \beta_i , v_i$) where mean
field approximation becomes exact. It would be a highly nontrivial task to 
investigate the physical properties of the hole regions of parameter space 
either
by infinite dimensional representations or by the explicit use of quadratic 
algebra. Consideration of $p$-species ASEP under fully parallel updating is 
another interesting subject which is under study.

\newpage
{ \large \bf Acknowledgement:} \\
We would like to thank V.Karimipour for his valuable comments and 
discussions. We also appreciate M.Abolhasani, A.Langari and J.Davoodi 
for their useful helps.


\end{document}